\documentclass[a4paper,11pt,twocolumn]{article}  
\pdfoutput=1
\usepackage[T1]{fontenc} 
\usepackage{graphicx} 
\usepackage{bbold}
\usepackage{mathtools}
\usepackage{amsmath}
\numberwithin{equation}{section} 
\usepackage{amsfonts}
\usepackage{amssymb} 
\usepackage{slashed} 
\usepackage{color}
\usepackage[table]{xcolor}
\usepackage{tablestyles} 
\usepackage{tabularx}
\usepackage{hyperref} 
\usepackage{bm}
\usepackage{cite}  
\usepackage{mathrsfs}
\usepackage{framed}
\usepackage[font=footnotesize,labelfont=bf]{caption}
\def\eq#1{{Eq.~(\ref{#1})}}
\def\eqs#1#2{{Eqs.~(\ref{#1})--(\ref{#2})}}

\def\Tr{\mbox{Tr}\,}

\colorlet{grayline}{gray!70}
\definecolor{blueline}{rgb}{0,0.27,0.55}
\definecolor{DarkGray}{gray}{0.4}
\definecolor{Gray}{gray}{0.6}
\definecolor{oucrimsonred}{rgb}{0.6, 0.0, 0.0}
\definecolor{persianblue}{rgb}{0.11, 0.22, 0.73}
\definecolor{forestgreen}{rgb}{0.13,0.35,0.13}
 \hypersetup{colorlinks, citecolor=forestgreen, linkcolor=forestgreen, urlcolor=forestgreen}
\newcommand{\be}{\begin{equation}}
\newcommand{\ee}{\end{equation}}
\newcommand{\bea}{\begin{eqnarray}}
\newcommand{\eea}{\end{eqnarray}}
\newcommand{\nn}{\nonumber}

\newcommand*\xbar[1]{%
  \hbox{\;%
    \vbox{%
      \hrule height 0.5pt 
      \kern0.5ex
      \hbox{%
        \kern-0.25em
        \ensuremath{#1}%
        \kern-0.07em
      }%
    }%
  }%
} 
\newcommand{\com}[1]{}
\newcommand{\gsim}{\lower.7ex\hbox{$\;\stackrel{\textstyle>}{\sim}\;$}}
\newcommand{\lsim}{\lower.7ex\hbox{$\;\stackrel{\textstyle<}{\sim}\;$}} 

\newcommand{\bc}{\begin{center}}
\newcommand{\ec}{\end{center}}

\newcommand{\lambdap}{\lambda^{\prime}}
\newcommand{\lambdaA}{\lambda_{1}}
\newcommand{\lambdaB}{\lambda_{2}}
\newcommand{\lambdaAp}{\lambda_1^{\prime}}
\newcommand{\lambdaBp}{\lambda_2^{\prime}}
\newcommand{\mup}{\mu^{\prime}}
\newcommand{\nup}{\nu^{\prime}}
\newcommand{\hH}{h}
\newcommand{\fH}{f}
\newcommand{\gH}{g}
\font\beeg=cmr17 scaled 1800
\newbox\ibox
\def\versal#1{\setbox\ibox=\hbox{{\beeg #1}~}%
	    \noindent\global\hangindent=\wd\ibox\global\hangafter-2%
	    \sc\smash{\llap {\lower 14pt \box\ibox}}}

\newcommand{\LVV}{a_{V}^{ 2}}
\newcommand{\LV}{a_{V}}
\newcommand{\LtVV}{\widetilde{a}_{V}^{2}}
\newcommand{\LtV}{\widetilde{a}_{V}}
\newcommand{\DenH}{\Phi_H}
\newcommand{\cmb}{\mathscr{C}_2}
\begin{document}
\onecolumn
\thispagestyle{empty}
\begin{center}
{ \Large \color{oucrimsonred} \textbf{ 
 Stringent bounds on $HWW$ and $HZZ$ anomalous couplings  \\[0.3cm] 
with quantum tomography at the LHC
}}

\vspace*{1.5cm}
{\color{DarkGray}
  {\bf M. Fabbrichesi$^{a}$,}
{\bf   R. Floreanini$^{a}$,}
{\bf E. Gabrielli$^{{b,a,c,d}}$ and} 
 {\bf L. Marzola$^{{d}}$}
}\\

\vspace{0.5cm}
{\small 
{\it  \color{DarkGray} (a)
INFN, Sezione di Trieste, Via Valerio 2, I-34127 Trieste, Italy}
\\[1mm]
  {\it \color{DarkGray}
    (b) Physics Department, University of Trieste, Strada Costiera 11, \\ I-34151 Trieste, Italy}
  \\[1mm]  
  {\it \color{DarkGray} 
   (c) CERN, Theoretical Physics Department, Geneva, Switzerland}
    \\[1mm]
  {\it \color{DarkGray}
(d) Laboratory of High-Energy and Computational Physics, NICPB, R\"avala 10, \\ 10143 Tallinn, Estonia}
}
\ec

 \vskip0.5cm
\bc
{\color{DarkGray}
\rule{0.7\textwidth}{0.5pt}}
\ec
\vskip1cm
\bc
{\bf ABSTRACT}
\ec

\vspace*{5mm}

\noindent 
Quantum tomography provides the full reconstruction of the density matrix of a state. We use it to study the Higgs boson decay into weak gauge bosons. Anomalous couplings beyond the Standard Model can be constrained by means of observables easily defined in terms of the polarization density matrix. We describe a strategy based on three observables that together provide the most stringent limits. Two of these observables are linked to the entanglement between the polarizations of the two gauge bosons, the other is based on CP-odd combinations of one momentum and two polarizations. We find for the $Z$ channel that this strategy offers, already with the available LHC data, limits competitive with the best  available bounds. We argue that the inclusion of these  observables in routine experimental analyses can lead to more stringent global fit limits.
  
  \vskip 3cm
\bc 
{\color{DarkGray} \vbox{$\bowtie$}}
\ec

\newpage
\section{Introduction\label{sec:intro}} 

{\versal The decay of the Higgs boson} provides an ideal laboratory for a systematic study  of weak gauge bosons polarizations. The massive  gauge  bosons---whose  polarizations  represent quantum states with three possible levels, that is, \textit{qutrits}---act as their own polarimeters and  the full polarization density matrix    can be reconstructed, within the inherent uncertainties of the procedure, from the angular distribution of the final leptons in what has been dubbed \textit{quantum tomography}.

This opportunity has been explored in a series of papers~\cite{Barr:2021zcp,Aguilar-Saavedra:2022wam,Ashby-Pickering:2022umy,Fabbrichesi:2023cev} in which  the polarization density matrix of the bipartite system formed by the two gauge bosons has been computed, and observables quantifying the \textit{entanglement}~\cite{Horodecki:2009zz} and the violation of the Bell inequalities~\cite{Bell} were analyzed.

In this work we utilize the polarization density matrix of the processes $H\to W  W^*$ and $H\to Z  Z^*$ (where $W^*$ and $Z^*$ denote off-shell states) to study the effect of possible anomalous couplings between the Higgs boson and the weak gauge bosons. The study of anomalous couplings is an important area of research in particle physics because their existence would imply the presence of new particles or interactions. Their precise measurement can then provide insights into the nature of new physics beyond the Standard Model (SM). We are particularly interested in the CP nature of these couplings and the related question of the parity of the Higgs boson, that is, whether it  has or not a pseudo-scalar component.

The most general Lagrangian for the processes we are interested in can be written as
\bea
{\cal L}_{HVV} & = & g\, M_{W} W^{+}_{\mu}W^{-\mu} H + \frac{g}{2 \cos\theta_{W}} M_{Z} Z_{\mu}Z^{\mu} H\nn \\
&& -\frac{g}{M_{W} }\Bigg[ \frac{a_{W}}{2} W^{+}_{\mu\nu} W^{-\mu\nu} +
\frac{\widetilde a_{W}}{2} W^{+}_{\mu\nu}\widetilde W^{-\mu\nu}  +\,\frac{a_{Z}}{4} Z_{\mu\nu} Z^{\mu\nu}  + 
\frac{\widetilde a_{Z}}{4} Z_{\mu\nu}\widetilde Z^{\mu\nu}\Bigg]  H\, ,\label{L}
\eea
The first line in \eq{L} gives the SM Lagrangian, the second introduces two possible anomalous couplings for each of the gauge bosons $V=W$ or $Z$: $a_{V}$ and $\widetilde a_{V}$. In \eq{L}, $g$ is the $SU(2)$ coupling, $V^{\mu\nu}$ are the field strength tensors and $ \widetilde V^{\mu\nu} = \frac{1}{2} \epsilon^{\mu\nu\rho\sigma} V_{\rho\sigma}$. All couplings in \eq{L} are taken to be real. The coupling $a_{V}$ stands for a departure of the fundamental interaction from that of the SM. A non-vanishing  coupling $\widetilde a_{V}$  signals the presence of a pseudo-scalar component in the Higgs boson and the possibility of CP violation in the interference with the SM vertex. 

Numerous works have studied  the anomalous couplings in \eq{L} by means of dedicated observables~\cite{Soni:1993jc,Chang:1993jy,Skjold:1993jd,Buszello:2002uu,Choi:2002jk,Gao:2010qx,Christensen:2010pf,Desai:2011yj,Bolognesi:2012mm,Dwivedi:2016xwm,Anderson:2013afp} and in the framework of effective field theories~\cite{Artoisenet:2013puc,Boselli:2017pef,Brivio:2019myy}---even though most of the references only consider observables that are combinations of the final lepton momenta and energies. The  structure of the helicity amplitudes for the considered processes has been investigated in~\cite{Hellmund:1981kq,Shim:1995ax,Mahlon:1998jd,Bern:2011ie,Stirling:2012zt,Maina:2020rgd,Maina:2021xpe} and applied to the anomalous couplings in~\cite{Rao:2020hel}.

In this work, we introduce a new strategy that exploits the full polarization density matrix.  Knowledge of the density matrix gives a bird's-eye view of the possible observables available for a given process. Some of them, like those linked to entanglement, are untested yet. Others are already known and utilized, like products of momenta and polarizations or the cross section---the simplest of them all.
 
For the present case of the Higgs boson decays, we define three observables in terms of the entries of the polarization density matrix to provide a new mean to constrain the anomalous couplings in \eq{L}: two observables are linked to entanglement in the spin correlations, the third is related to products of one momentum and two polarizations and it is specific to the CP-odd vertex.  This combination of observables seems as sensitive or even more to any changes in the Lagrangian as analyses based on data from multiple cross sections and, 
therefore, it represents a useful tool to further build on this more traditional approach. 
 
Polarizations are more difficult to measure than momenta. The reconstruction of the polarization density matrix from the data is challenging, in particular   for the case of $H\to W  W^*$ because of the presence of the undetectable  neutrinos. The main aim of this work is to show that the extra work needed to reconstruct the gauge boson polarizations is indeed worthwhile. In order to present an analysis as realistic as possible, we give an estimate of the main uncertainties in the bounds that we derive for the anomalous couplings. In particular, we include the effects of statistical and systematic errors, as well as the impact of the dominant irreducible backgrounds in the computation of these observables. It is understood that a complete treatment of uncertainties goes beyond the purpose of the present work and should be the focus of a dedicated analysis that can only be performed by the experimental collaborations.
 
To assess how our method fares, we compare the bounds we derive with the theoretical analysis presented in~\cite{Rao:2020hel}. Comparison with experimental limits requires the inclusion of the irreducible background. We estimate this for the $ZZ^*$ channel, which achieves a higher sensitivity, and confront our findings with the limits due to a combination of cross sections recently updated the CMS~\cite{CMS:2019ekd} collaboration.

\section{Methods} 

{\versal Quantum tomography} aims to fully determine the density matrix $\rho$ of a quantum state. In the case of the decay of the Higgs boson into two massive spin 1 gauge bosons, the density matrix for the joint polarization states of the two particles is the $9\times 9$ matrix of a system composed by two qutrits. It can be decomposed on the basis formed by the eight Gell-Mann matrices $T^a$ and the unit matrix as follows
\be
\rho_H = \frac{1}{9}\left[\mathbb{1}\otimes
    \mathbb{1}\right]+   \sum_a f_a \left[T^a\otimes \mathbb{1}\right]  +
\sum_a g_a \left[\mathbb{1}\otimes T^a\right] 
    +\sum_{ab} h_{ab}  \left[T^a\otimes T^b\right]
\label{rho}
\ee
where the $T^a$ satisfy the orthogonality condition $\Tr[T^a T^b]= 2\, \delta^{ab}$ and the indices $\lambda$ and $\lambda^\prime$ track the polarizations of the two spin 1 massive particles. \eq{rho} defines the coefficients $f_a$, $g_a$, and $h_{ab}$ ($a,b\in\{1, \dots,8\}$) which we determine in what follows. 

To best constrain the anomalous couplings in the Lagrangian \eq{L}, we need observables that ideally depend linearly on these quantities. Quantum tomography gives the coefficients $f_a$, $g_a$, and $h_{ab}$ of the density matrix and there is a number of observables that can be constructed with them. Some of these observables are novel and based on the entanglement between the final states, others are linked to correlations already studied or, directly, to the cross section of the process. We consider the three that provide the most stringent limits to the anomalous couplings: 
\begin{itemize}
\item
When a scalar particle decays into a pair of massive gauge bosons, the qutrits describing the polarizations of the latter form a bipartite pure state. In this case it is possible to quantify the entanglement among the polarizations of the weak gauge bosons by computing the \emph{entropy of entanglement},   $\mathscr{E}_{ent}$, defined by
\be
\mathscr{E}_{{ent}} = - \Tr[\rho_A \log \rho_A] =  - \Tr[\rho_B \log \rho_B] \, , \label{E}
\ee
where $\rho_A$ and $\rho_B$ are the reduced density matrix obtained by tracing out the subsystem $B$ or $A$, respectively. In our case, the two subsystems are the polarizations of the two massive gauge bosons and the reduced density matrices are obtained by tracing out the polarizations of either gauge boson. For a two-qutrit system the entropy of entanglement satisfies $0\leq \mathscr{E}[\rho] \leq \ln 3$. The first equality is true if and only if the bipartite state is separable (signalling the absence of entanglement), the second if the bipartite state is maximally entangled;

\item The second observable $\cmb$~\cite{Mintert} is an entanglement witness defined in terms of the coefficients in \eq{rho} as
\bea
\cmb &=& 2\max \Big[0,\, -\frac{2}{9}-12 \sum_a f_{a}^{2} +6 \sum_a g_{a}^{2} + 4 \sum_{ab} h_{ab}^{2}\, ,\nn \Big.\\
& & \Big. -\frac{2}{9}-12  \sum_a g_{a}^{2} +6 \sum_a f_{a}^{2} + 4 \sum_{ab} h_{ab}^{2} \Big]\,.
\eea
It represents a lower bound on the \textit{concurrence}~\cite{Rungta} of the system, which quantifies the entanglement of a generic bipartite system. Unfortunately, the definition of concurrence is based on an optimization problem  that makes its evaluation  a very hard mathematical task---with a simple analytic solution only when the two subsystems  are two-level systems (qubits). For this reason, we use $\cmb$ as proxy for the concurrence.
Whereas the entropy of entanglement can be used to quantify the entanglement content only of pure states, the observable $\cmb$ can be utilized for arbitrary states. $ \cmb$ is at most equal to 4/3. 

\item
The third  observable $\mathscr{C}_{odd}$ singles out the asymmetric parts of the density matrix; it defined as
\be
\mathscr{C}_{odd}=\frac{1}{2}\, \sum_{\substack{a,b\\ a< b}} \Big| h_{ab} -h_{ba} \Big| \, , \label{CPodd} 
\ee
which contains only off-diagonal terms that change sign under transposition.  In the difference symmetric terms drop out. 
The combinations in \eq{CPodd} provide correlations. They can  be written in terms of the kinematical variables as triple products of momenta and polarizations as, for instance,
\be
\vec k \cdot \Big(\vec \varepsilon_{\hat n} \times \vec \varepsilon_{\hat r} \Big) \, , \label{Todd}
\ee
with $\vec k$ the momentum of one of the particles, $\vec \varepsilon_{\hat n}$ and  $\vec \varepsilon_{\hat r}$ the projections of the polarizations along two directions orthogonal to the momentum.

\end{itemize}

The decay of the Higgs boson shows at colliders  in the  processes 
\be
p\;p \to V_1 + V_2 + X \to \ell^+\ell^- \; (\text{or}\; \,\ell^+\, j j\; \text{or}\; \,\ell^-\, jj) + E^{\text{miss}}_{T} \; (\text{or}\; \,\ell^+\ell^-) + \text{jets} \, ,
\ee
with missing energy $E^{\text{miss}}_{T}$ due to the possible presence of neutrinos in the final state. These processes include the production of the gauge bosons through  the resonant Higgs boson channel, as well as via quark fusion, and include the consequent decays into the final leptons or quarks---plus the jets originating from the spectator quarks. 

The production of the Higgs boson is dominated  at the LHC by gluon fusion. This is the mode we study. Sub-leading, but not negligible, electroweak processes like vector boson fusion (VBF) contribute as well. In these processes the anomalous couplings enter off-shell. Some of the new physics vertices are enhanced by the energy and make some of these off-shell contribution more sensitive. Moreover, they may enter twice, first in production and then in the decays. For these reasons, comparison of bounds obtained from different processes as well as a direct comparison of our bounds to others is often not straightforward.

\begin{figure}[h!]
\begin{center}
\includegraphics[width=3in]{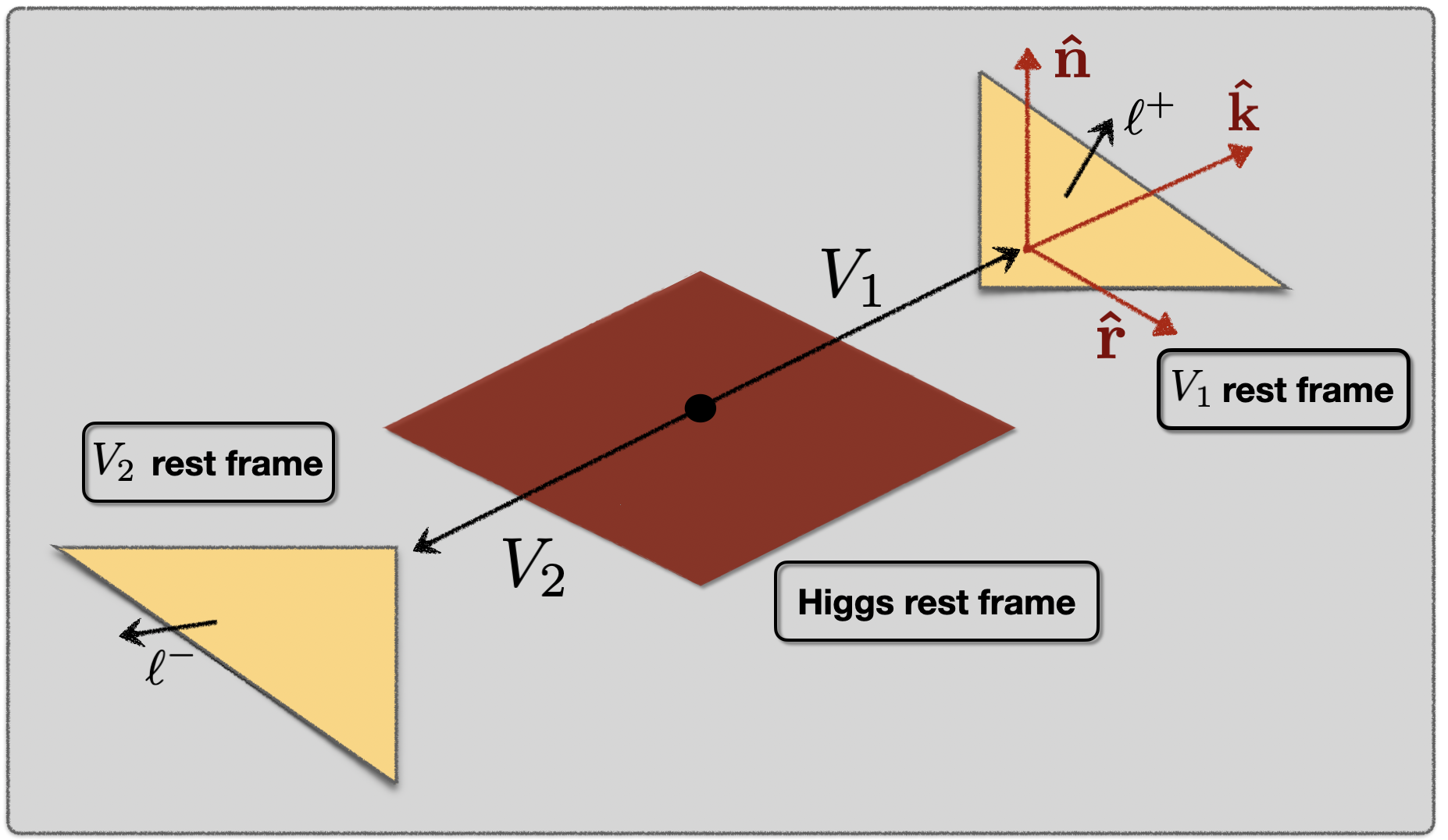}
\caption{\small Unit vectors and momenta utilized in the text to describe the decay of the Higgs boson into the weak gauge bosons $V_1$ and $V_2$. 
\label{fig:coordinates} 
}
\end{center}
\end{figure}

The spin 1 gauge bosons act as their own polarimeters and the momenta of the final leptons (see Fig.\ref{fig:coordinates}) provide a measurement of the gauge boson polarizations. The quantum tomography problem of reconstructing the correlation coefficients $h_{ab}$, $f_a$ and $g_a$ from the final states of the process at hand has been recently discussed in~\cite{Ashby-Pickering:2022umy}, to which we refer for more details. 

In the case of the Higgs boson decay, the computation of the correlation coefficients  is greatly simplified because there are only two independent entries in the density matrix that need to be determined. All the elements of the density matrix  can be written either in terms of the Gell-Mann basis or of that of the helicity amplitudes:
\be
\label{helamp}
h_{\lambda} = \langle V(\lambda) V^{*}(-\lambda)| -{\cal L} |H\rangle\quad \text{with} \quad \lambda=0,\, \pm\,,
\ee
with ${\cal L}$ the Lagrangian given in \eq{L}. Helicities are here defined with respect to the $\hat z$ direction  in the rest frame of the first spin-1 particle.

\subsection{Estimating the uncertainty}

 In order to evaluate the sensitivity of the experiments to the anomalous couplings in the observables $\mathscr{E}_{ent}$,  $\cmb$ and $\mathscr{C}_{odd}$,  we first estimate the number of suitable events which are available.
  
The cross sections (at $\sqrt{s}=13$ TeV) of the two processes we are interested in are
\be
 \sigma(p \,p \to H\to W^+\ell^-\bar \nu_\ell) = 12.0 \pm 1.4 \;\; \text{pb~\cite{ATLAS:2022ooq}} \label{pb1}
 \ee
  and
  \be
  \sigma(p \,p \to H\to  Z\ell^+\ell^-)= 1.34\pm 0. 12 \;\; \text{pb~\cite{ATLAS:2020rej}.}\label{pb2}
\ee
These numbers must be multiplied by the corresponding branching ratios. Among the leptonic ones, we only retain those into electrons and muons. In both cases we have a sufficient number of events to compute the polarizations of the gauge bosons, of the order of $10^5$ at the LHC run2 for the $WW^*$ process and roughly $10^3$ for the $ZZ^*$ process.
 
As it is often the case, the definition of better observables from the  theoretical point of view goes hand in  hand with a more challenging reconstruction of the same from the data.

We take into account the problem of the irreducible uncertainties  in the evaluation of the operators for the decays of the $WW^*$  by considering  the semi-leptonic decay $H\to jj\ell \nu_\ell$ (rather than the fully leptonic one) and use the momentum from the $s$-jet (identified via the $c$-tagging of the companion jet)  to measure the polarization of one of the two $W$-bosons. It has been shown that the efficiency of the jet tagging and the decreased uncertainty in the single neutrino momentum may improve the polarization reconstruction~\cite{maurin}.

The uncertainty on each of the observables considered, let us call them  $O(a_{V},\widetilde a_{V})$, is found by the following procedure. The main source of uncertainty comes from the determination of the polarizations and it originates in the reconstruction of the rest frame of the decaying Higgs boson. The same reconstruction is also necessary in the determination of the Higgs boson mass $m_H$ from the events in which the Higgs boson decays into either $ZZ^*$ or $WW^*$; we therefore can use the related reconstruction error as a proxy for the dominant uncertainty in our computations. We propagate the uncertainty on the value of $m_H$ to the operators by means of a Monte Carlo simulation, obtaining the related variances $\sigma_i^2$ for each of the considered decays as $m_H$ is randomly varied within its experimental uncertainty. This uncertainty includes both statistical and systematic errors.

The procedure works well for the $ZZ^*$ case where it is found that $m_H=124\pm0.18\pm0.04$\cite{ATLAS:2022zjg}. For the $WW^*$ channel,  only  the transverse mass can be determined, and that comes with an error (1$\sigma$) of about 5 GeV for the fully leptonic decays~\cite{CMS:2022uhn}. We make here an educated guess by taking half of this uncertainty in the case of the semi-leptonic decays (for which the transverse mass is in addition closer to $m_H$).

To constrain the values of the anomalous couplings $a_V$ and $\widetilde a_V$ we introduce a $\chi^2$ test set for a 95\% CL
\be
\sum_{i} \left[ \frac{O_i(a_{V},\widetilde a_{V})-O_i(0,0)}{\sigma_{i}}\right]^2 \leq 5.991 \,,  \label{chi2}
\ee
in which we set the uncertainties $\sigma_i$ as discussed above. The choice of operators utilized in the test depends on the purity of the $WW^*$ and $ZZ^*$ states as clarified below.

\section{Results}
{\versal Consider the decay}
\be
H\to V(k_1,\lambda_1)\, V^*(k_2,\lambda_2)\, ,
\label{HVV}
\ee
with $V$ either $W$ or $Z$, and $V^*$ regarded as an off-shell vector boson. Here $k_{1}$ and $k_2$ denote the associated particle momenta and the helicities $\lambda_{1,2}$ take values  $\lambda_{1,2}=\{+1,0,-1\}$.

In the following, we treat the latter as an on-shell particle characterized by a mass 
\be
M^*_V= f M_V
\ee
reduced by a factor $0<f<1$ with respect to the original mass $M_V$.
\begin{figure}[h!]
\begin{center}
\includegraphics[width=3.5in]{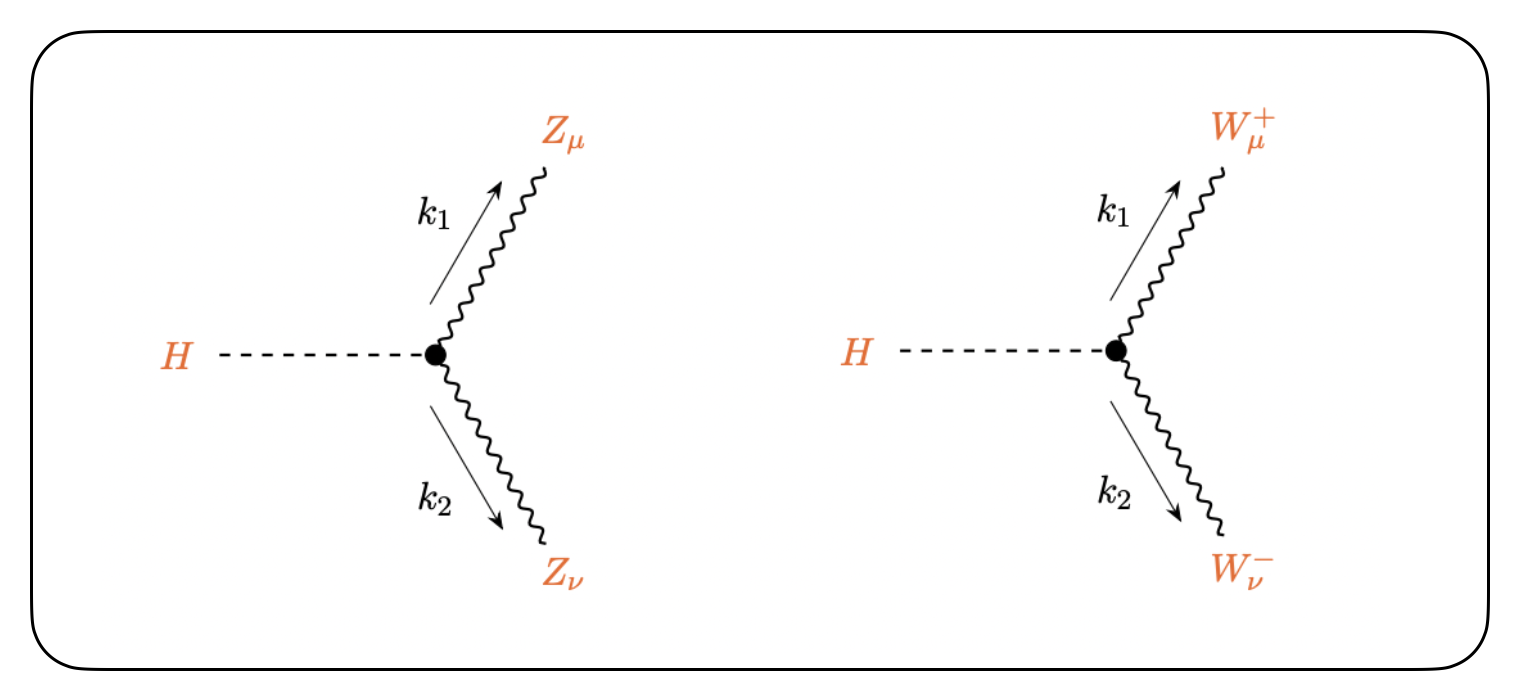}
\caption{\small Feynman diagrams for the decay of the Higgs boson into a pair of gauge bosons, the vertex includes the anomalous couplings in \eq{L}.
\label{fig:HtoVV} 
}
\end{center}
\end{figure}
From the Lagrangian in \eq{L}, the amplitude of the Higgs boson decay (\ref{HVV}) is given by 
\bea
{\cal M}(\lambdaA,\lambdaB)= M_{\mu\nu}
\varepsilon^{\mu\star}(k_1,\lambdaA)\varepsilon^{\nu\star}(k_2,\lambdaB)
\, ,
\label{MHVV}
\eea
with
\bea
M^{\mu\nu}=g\, M_V\xi_V \, g^{\mu\nu} -\frac{g}{M_W}
\Big[a_V \left(k_1^{\nu} k_2^{\mu}-g^{\mu\nu} k_1\cdot k_2\right)+
\tilde{a}_V \epsilon^{\mu\nu\alpha\beta}k_{1\alpha} k_{2\beta}\Big]\, ,  
\eea
where $g$ is the weak coupling, $\xi_W=1$, and $\xi_Z=1/(2c_W)$, with $c_W=\cos{\theta_W}$ and $\theta_W$ the Weinberg angle. From the amplitude in Eq.~(\ref{MHVV}) we obtain
\bea
{\cal M}(\lambdaA,\lambdaB) {\cal M}(\lambdaAp,\lambdaBp)^{\dag} =
M_{\mu\nu}M^{\dag}_{\mup\nup}  \mathscr{P}^{\mu\mup}_{\lambdaA\lambdaAp}(k_1)
 \mathscr{P}^{\nu\nup}_{\lambdaB\lambdaBp}(k_2)\, . \label{Mpol}
\eea
where $\mathscr{P}^{\mu\nu}_{\lambda\lambda^{\prime}}(k)$ is equal to~\cite{Song:1979wc,Choi:1989yf} 
\bea
\mathscr{P}^{\mu\nu}_{\lambda \lambdap}(p) &=&\varepsilon^{\mu}(p,\lambda)^{\star}\varepsilon^{\nu}(p,\lambdap) \nn\\&=&
\frac{1}{3}\left(-g^{\mu\nu}+\frac{p^{\mu}p^{\nu}}{M^2_V}\right)
\delta_{\lambda\lambdap}-\frac{i}{2M_V}
\epsilon^{\mu\nu\alpha\beta}p_{\alpha} n_{\beta}^i \left(S_i\right)_{\lambda\lambdap}-\frac{1}{2}n_i^{\mu}n_j^{\nu} \left(S_{ij}\right)_{\lambda\lambdap}\, ,
\label{proj}
\eea
with $S_i$, $i\in\{1,2,3\}$, being the spin-1 matrix representations of the $SU(2)$ generators. The matrices $S_{ij}$ are defined as
$
S_{ij}= S_iS_j+S_jS_i-\frac{4}{3} \mathbb{1}\, \delta_{ij} \, ,
$
with $i,j\in\{1,2,3\}$ and $\mathbb{1}$ being the $3\times 3$ unit matrix.  \eq{proj} with $M=M_V$ or $M=M^*_V$ for the on-shell and off-shell boson, respectively.\footnote{  
For the helicity basis  we use a representation
  where the eigenstates of $S_3$ read 
\begin{equation}
  |+\rangle = 
  \begin{pmatrix}
  1 \\ 0 \\ 0  
  \end{pmatrix},
  \quad
  |0\rangle = 
  \begin{pmatrix}
  0 \\ 1 \\ 0  
  \end{pmatrix},
  \quad
  |-\rangle = 
  \begin{pmatrix}
  0 \\ 0 \\ 1  
  \end{pmatrix}\,,
\label{hbasis}
\end{equation}
corresponding to the eigenvalues $\lambda=+1,0,-1$, respectively, and
the symbols $\left(S_i\right)_{\lambda\lambdap}$ and $\left(S_{ij}\right)_{\lambda\lambdap}$ are the corresponding matrix elements of the matrices $S_i$ and $S_{ij}$ on this basis respectively.}

 By using the expression in Eq.~(\ref{Mpol}) we have that
\bea
\rho_H =\frac{{\cal M}_{\mu\nu}{\cal M}^{\dag}_{\mu^{\prime}\nu^{\prime}}}{|\xbar{{\cal M}}|^2} \Big[\mathscr{P}^{\mu\mu'}(k_1)\otimes
\mathscr{P}^{\nu\nu'}(k_2)\Big]
  \, , \label{xxx}
\eea
where the expression for $\mathscr{P}^{\nu\nu'}(k_{1,2})$ is given in Eq.~(\ref{proj}) and $|\xbar{{\cal M}}|^2$ stands for the unpolarized square amplitude of the process, that for $V=W$ reads
\bea
|\xbar{{\cal M}}|^2&=&\frac{g^2}{4 f^2 M_V^2}\DenH\, ,
\eea
which in turn gives the cross section. The coefficients $f_a,g_a,h_{ab}$ in \eq{rho} can be obtained from the relation in \eq{xxx} upon a projection of the spin matrices $S_i$ and their products on the Gell-Mann basis \cite{Fabbrichesi:2023cev}. The matrix $\rho_H$ above satisfies the unitarity relation $\Tr[\rho_H]=1$.

The relation in \eq{xxx} provides a simple way to compute the polarization density matrix of the  massive spin-1 particles starting from the amplitudes  of the related production process. 
In the case of $V=W$, we find that the non-vanishing $f_{a}$ elements are given as
\bea
f_{3}&=&-\frac{1}{6 \DenH}\,
\Big[1  - 
  f^2 \Big(\LtVV + \LVV  \Big)\Big]\Delta_{H}
\, ,
\nonumber\\
f_{8}&=&-\frac{1}{\sqrt{3}}f_3\, ,
\label{faHiggs}
\eea
with $g_{a}=f_a$ for $a\in\{1,\dots,8\}$, and where
\bea
\DenH&=&\Big[ 1  + 2\, f^2
  \left(\LtVV + \LVV \right)\Big] m_H^4
-2 \Big[1  + 
  f^2 \Big(1 + 2 \LtVV + 2 \LVV  - 6 \LV \Big)
\nonumber\\
&+&   2\,  f^4 \Big(\LtVV + \LVV \Big) m_H^2 M_V^2+\Bigg[1  + 
  2 f^6 \Big(\LtVV + \LVV \Big)
\nonumber\\
&+&    2 f^2 \Big(5 + \LtVV + \LVV  - 6 \LV \Big) +    f^4 \Big(1 - 4 \LtVV + 8 \LVV - 
      12 \LV\Big)\Bigg] M_V^4
\eea
and
\be
\Delta_{H} = \Big[m_H^4 - 2 \left(1 + f^2\right) m_H^2 M_V^2 +
  \left(1 - f^2\right)^2 M_V^4\Big]\, .
\ee

The non-vanishing $ h_{ab}=\tilde{h}_{ab}/\DenH$ elements for $V=W$ are  given as
\bea
\tilde{h}_{16}&=&
-\frac{1}{2}\, f \Big\{
  \LV m_H^2 - \Big[\left(1 + f^2\right) \LV 
  - 2 \Big]M_V^2\Big\}
\, \Big\{ m_H^2 -
\Big[1 +
     f^2 \left(1 - 2 \LV \right)\Big] M_V^2\Big\}\, ,
    \nonumber\\
    \tilde{h}_{61}&=&\tilde{h}_{16}=\tilde{h}_{27}=\tilde{h}_{72}\, ,
    \nonumber\\
    \tilde{h}_{17}&=&-\frac{1}{2} f \, \LtV \, 
    \Big\{ m_H^2 
    - \Big[1 +
    f^2 \left(1 - 2 \LV \right)\Big] M_V^2\Big\}\sqrt{\Delta_H}\, ,
    \nonumber\\
    \tilde{h}_{71}&=&-\tilde{h}_{17}= \tilde{h}_{26}=-\tilde{h}_{62}\, ,
    \nonumber\\
    \tilde{h}_{45}&=&
f^2 \, \LtV \, \Big\{\LV m_H^2 + \Big[2  - (1 + f^2) \LV \Big] M_V^2\Big\} \sqrt{\Delta_H}\, ,
    \nonumber\\
    \tilde{h}_{54}&=&-\tilde{h}_{45}\, , \nn \\
    \tilde{h}_{33}&=&\frac{1}{4}\,
    \Big\{ m_H^2 - \Big[1  + 
      f^2 \left(1 - 2 \LV \right)\Big] M_V^2\Big\}^2\, ,
    \nonumber\\
\tilde{h}_{38}&=&\tilde{h}_{83}=-\frac{1}{4\sqrt{3}}\DenH\, ,
\label{hHiggs}
\\
    \tilde{h}_{44}&=&\tilde{h}_{55}=
    \frac{1}{2}\, f^2 \Big\{\Big[\LV m_H^2 + 2 M_V^2 - (1 + f^2) \LV
      M_V^2\Big]^2-\LtVV \,\Delta_H\Big\}\, ,
\nonumber\\
\tilde{h}_{88}&=&\frac{1}{12}\, \Big\{\Big[ 1
- 4\, f^2 (\LtVV + \LVV ) \Big]m_H^4 -  2\, \Big[1  + 
      f^2 (1 - 4 \LtVV - 4 \LVV + 
      6 \LV \big)
   - 4\, f^4 (\LtVV + \LVV ) \Big] m_H^2 M_V^2\nonumber\\
&&
+ \Big[1  - 
      2\, f^2 (7 + 2 \LtVV + 2 \LVV - 
        6 \LV)
  +f^4 (1 + 8 \LtVV - 4 \LVV  + 
  12 \LV )
  -  4 f^6 ( \LtVV + 2\LVV ) \Big] M_V^4 \Big\} \, .\nn
\eea
The dependence of the polarization entanglement on the  mass of the virtual state  is due to the contribution of the longitudinal polarization. At  threshold it yields a  singlet state and the maximum of entanglement.
  Analogous results hold in the case of the $V=Z$ case, with the appropriate
  replacement of the anomalous couplings and SM coefficient $\xi_Z$.
 
The density matrix for the Higgs decay, as embodied by the coefficients in
\eqs{faHiggs}{hHiggs}, describes a pure state, that is $\rho_H^2 = \rho_H$, in the SM~\cite{Aguilar-Saavedra:2022wam,Fabbrichesi:2023cev} and after adding the anomalous couplings as well.  This remarkable fact follows from the state being, so to speak, prepared by the formation of the spin-0 decaying particle which turns a generic mixed state (as that produced by  colliding protons) into a pure one.
This state can be written as 
\be
|\Psi \rangle = \frac{1}{|\xbar{{\cal M}}|} \Big[  h_+\, |V(+)  V^*(-)\rangle + h_0 \, |V(0)  V^*(0)\rangle+  h_-\, |V(-)  V^{*}(+)\rangle \Big] \, ,\label{pure}
\ee
with
\be
|\xbar{{\cal M}}|^2= |h_0|^2 + |h_+|^2 + |h_-|^2 \, ,
\ee
where the helicity amplitudes $h_{\lambda}$  are defined in \eq{helamp}.

The relative weight of the transverse components $ |V(+)  V^{*}(-)\rangle $ and $ |V(-)  V^{*}(+)\rangle $ with respect to the longitudinal one $| V(0)  V^{*}(0 )\rangle$ is controlled  by the conservation of angular momentum. In general, only the helicity is conserved and the state in \eq{pure} belongs to the $J_z=0$ component of either the $J=0$, $1$ or $2$ states or a linear combination of them. For the SM model Higgs, for which $h_-=h_+$, the pure state in \eq{pure} is given by 
\be
|\Psi \rangle = \frac{1}{\sqrt{2 +  \varkappa^2}} \Big[  |V(+)  V^*(-)\rangle -  \varkappa \, |V(0)  V^*(0)\rangle+   |V(-)  V^{*}(+)\rangle \Big] \, .\label{pure2}
\ee
with $\varkappa=1+(m_H^2-(1+f)^2M_V^2)/(2fM_V^2)$~\cite{Fabbrichesi:2023cev}. The state in \eq{pure2} is the singlet state when $\varkappa=1$---which happens if the final vector bosons are produced at rest.

After making the Kronecker product in \eq{xxx} explicit, the resulting $9\times 9$ polarization density matrix $\rho = |\Psi \rangle \langle \Psi |$ is written as
\be
\small
\rho_{H} =  \frac{1}{|\xbar{{\cal M}}|^2}\, \begin{pmatrix} 
  0 & 0 & 0 & 0 & 0 & 0 & 0 & 0 & 0  \\
  0 & 0 & 0 & 0 & 0 & 0 & 0 & 0 & 0  \\
  0 & 0 &  h_+ h_+^* & 0 &  h_+  h_0^*& 0 &  h_+ h_-^*& 0 & 0  \\
  0 & 0 & 0 & 0 & 0 & 0 & 0 & 0 & 0  \\
  0 & 0 & h_0 h_+^* & 0 & h_0  h_0^*  & 0 & h_0 h_-^*& 0 & 0  \\
  0 & 0 & 0 & 0 & 0 & 0 & 0 & 0 & 0  \\
  0 & 0 &  h_- h _+^*& 0 &  h_- h_0^*& 0 &   h_-h_-^*& 0 & 0  \\
  0 & 0 & 0 & 0 & 0 & 0 & 0 & 0 & 0  \\
  0 & 0 & 0 & 0 & 0 & 0 & 0 & 0 & 0  \\
\end{pmatrix} \, ,
\label{rhoBVV}
\ee
in which, in terms of the anomalous couplings in \eq{L}, we have
 \bea
 h_0 &=&-A x - B(x^2-1)\, ,\nn \\
 h_{\pm}&=& A\mp C\sqrt{x^2-1}\, .
 \label{amph}
\eea
The coefficients $A$ and $B$ for $V=W$ are given by
\bea
A&=&g\left(M_V+a_V \frac{k_1\cdot k_2}{M_V}\right)\nn \\
B&=&-g\,a_V\,M_V\, ,~~~~C\,=\,ig\,\tilde{a}_V\,M_V
\eea
with $x = m_H^2/(2 f M_{V}^{2}) - (f^2+1)/(2f)$.
The amplitudes entering the density matrix in \eq{rhoBVV} can be written in terms of the $f_a,g_a,h_{ab}$ coefficients in the Gell-Mann basis as
\bea
\hat{h}_- \hat{h}_-^* &=& \frac{1}{9}\Big[1+3\sqrt{3}~ \big(\fH_8-2 \gH_8 -2\hH_{38}\big)
+ 9 \fH_3  - 6 \hH_{88}\Big]  \, , \nn \\
\hat{h}_0 \hat{h}_-^* &=& \hH_{16} + i~\big( \hH_{17} -  \hH_{26}\big) + \hH_{27} \, , \nn \\
\hat{h}_+ \hat{h}_-^*  &=& \hH_{44} + i~\big( \hH_{45} - \hH_{54}\big) + \hH_{55} \, , \nn \\
\hat{h}_0 \hat{h}_0^* &=& \frac{1}{9}\Big[1 - 9 \big(\fH_3 +  \gH_3 -\hH_{33}\big)
+ 3\sqrt{3}~ \big(\fH_8 +  \gH_8
- \hH_{38} - \hH_{83}\big) + 3 \hH_{88}\Big] \, , \nn \\
\hat{h}_+ \hat{h}_0^* &=& \hH_{61} + i~ \big(\hH_{62} - \hH_{71}\big) + \hH_{72} \, , \nn \\
\hat{h}_+ \hat{h}_+^* &=& \frac{1}{9}\Big[1+3\sqrt{3}~ \big(\gH_8-2 \fH_8 -2\hH_{83}\big)
+ 9 \gH_3  - 6 \hH_{88}\Big] \, ,
\label{gmh}
\eea
where $\hat{h}_{\lambda}\equiv h_{\lambda}/|\xbar{{\cal M}}|$.
\eq{gmh}  makes it possible to go from the Gell-Mann basis to that of the helicities. In the limit of vanishing anomalous couplings, the  results in
Eqs.(\ref{faHiggs})-(\ref{hHiggs})  go into those corresponding to the SM~\cite{Fabbrichesi:2023cev}.

The main theoretical uncertainty affecting the correlation coefficients in \eqs{faHiggs}{hHiggs} is due to higher order corrections to the tree-level values. To estimate the size of these contributions, we take as guidance the results in~\cite{Actis:2009ti,Boselli:2015aha}---in which the NLO corrections have been computed---and assume that the error induced by these missing corrections yields approximately 1\% of uncertainty on the main entanglement observables in the relevant kinematic regions.

Having determined the entries in the density matrix, we now study the dependencies of the observable introduced in the previous Section on the anomalous couplings.  
\begin{itemize}
\item[-] As mentioned before, the decays of the Higgs boson lead to the production of bipartite polarization states that are pure. When neglecting the background processes we can then use the entropy of entanglement to measure the entanglement among the polarizations of the massive gauge bosons.  Only  $\LV$ enters linearly in $\mathscr{E}_{ent}$; the coupling $\LtV$ by itself only enters quadratically. We do not write out the explicit expression of this operator because it is cumbersome (involving, as it does,  the eigenvalues necessary in the definition of the logarithm of a matrix); 
 
 \item[-] The leading dependence of the  observable $\mathscr{C}_{odd}$  is linear in $\LtV$. The anomalous couplings $\LV$ only enters in $\mathscr{C}_{odd}$ multiplied by $\LtV$ and is therefore quadratically suppressed. Utilizing the coefficients in \eqs{faHiggs}{hHiggs}, the observable is given by 
\be
\mathscr{C}_{odd}=
\frac{\LtV f (1 + \LV f)\sqrt{\Delta_H}}{\Phi_H}\Big\{ m_H^2 + (1 - f)^2 M_V^2\Big\}\, .
\ee
\end{itemize} 

\begin{figure}[h!]
\begin{center}
\includegraphics[width=3in]{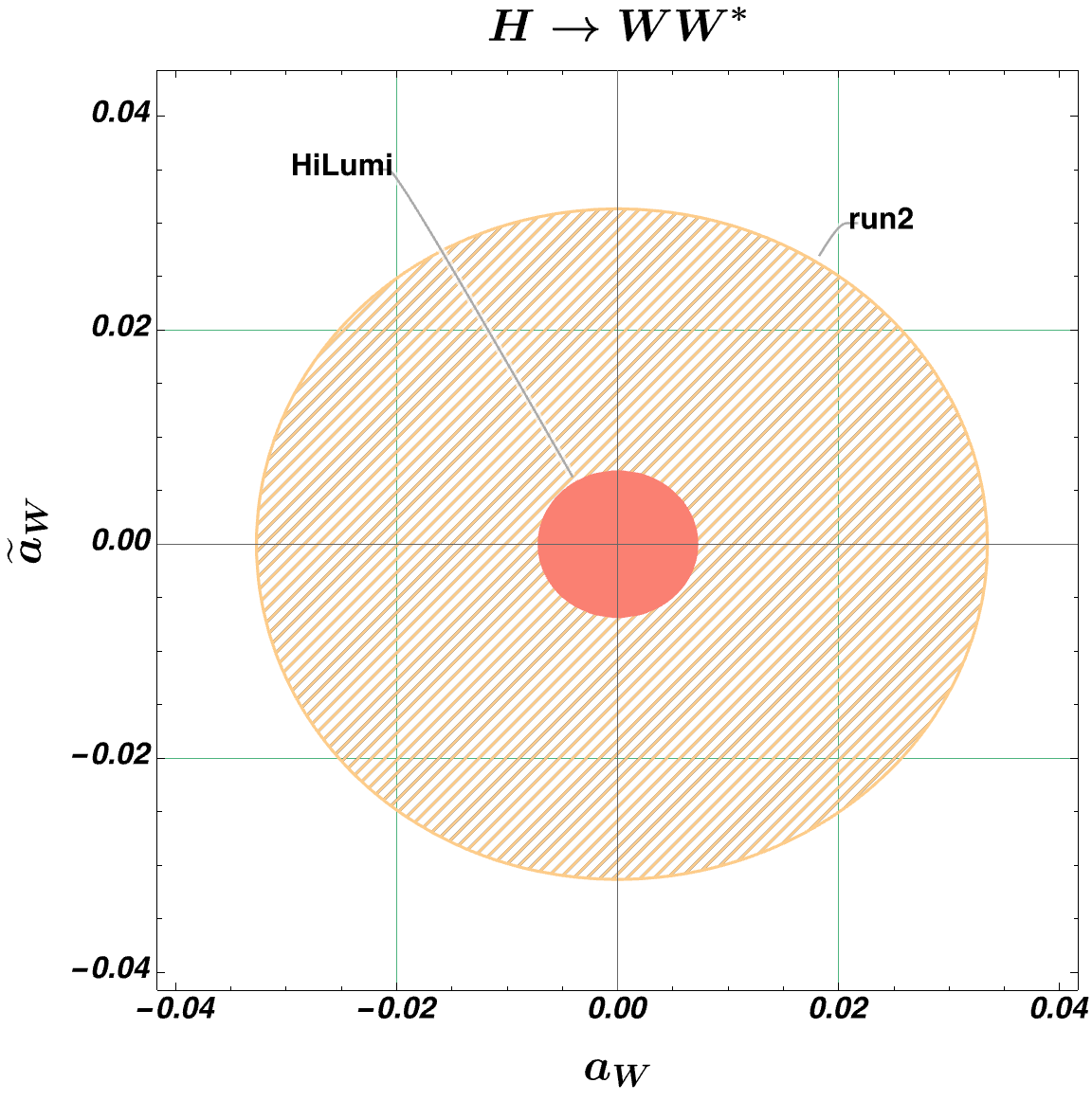}
\hspace{.5cm}
\includegraphics[width=3in]{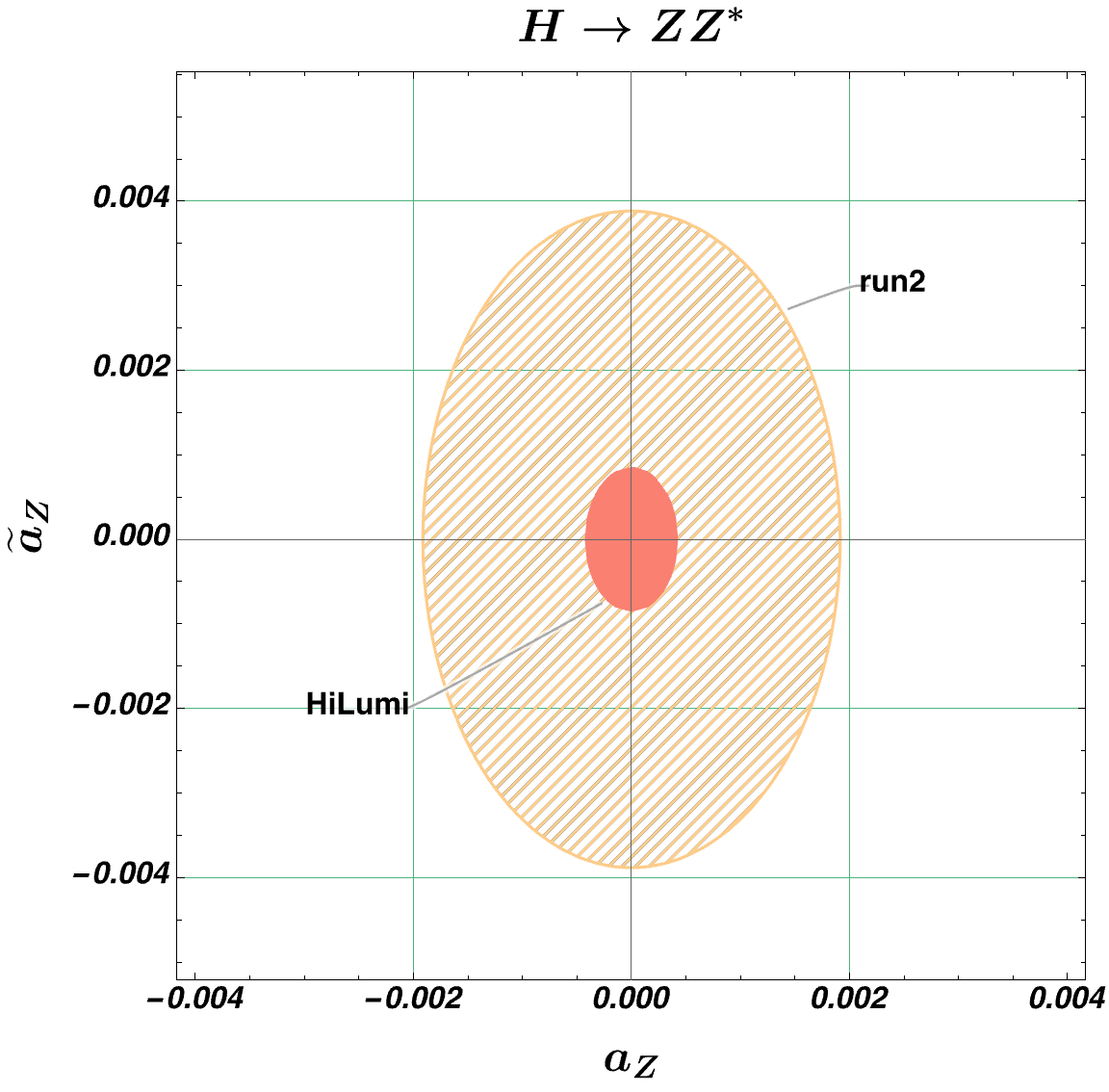}
\caption{\footnotesize Allowed values for the anomalous couplings $a_V$ and $\widetilde a_{V}$ obtained by using the observables $\mathscr{C}_{odd}$ and $\mathscr{E}_{ent}$. The hatched area use the LHC run2 data (${\cal L} = 140\; \text{fb}^{-1}$), the purple ones show the HiLumi projection (${\cal L} = 3\; \text{ab}^{-1}$). The limits, all given at a 95\% confidence level, only hold prior to the inclusion of backgrounds.
\label{fig:limits1} 
}
\end{center}
\end{figure}

The two observables $\mathscr{E}_{ent}$ and $\mathscr{C}_{odd}$ seem to be ideal inasmuch as  each of them depends   linearly on one of the anomalous couplings while marginalizing the other. Cross dependencies are quadratic and  very small in the range of values we consider. 

The two observables depend on the mass of the off-shell gauge bosons, parametrized  by $f M_W$. We have taken their average value by integrating over the parameter $f$ within its limits to obtain the results shown in Fig.~\ref{fig:limits1}. The limits for the Hi-Lumi case are obtained by rescaling the uncertainties by the square root of the ratio of the luminosities $\sqrt{300/14}$. The limits on single anomalous couplings, presented in Tab.~\ref{tab:thcouplings}, are obtained by marginalization of the other one. 

\begin{table}[t]
  \tablestyle[sansboldbw]
  \begin{tabular}{*{3}{p{0.2\textwidth}}}
  \theadstart
      \thead LHC &\thead  run2&
      \thead  HiLumi  \\
  \tbody
  & $ |a_{W}| \leq 0.033 $   & $|a_{W}|\leq 0.0070$ \\
  &   $|\widetilde a_{W}|\leq 0.031$   & $|\widetilde a_{W}|\leq 0.0068$  \\
  & $ |a_{Z}| \leq 0.0019 $   & $|a_{Z}|\leq 0.00040$ \\
  &   $|\widetilde a_{Z}|\leq 0.0039$   & $|\widetilde a_{Z}|\leq 0.00086$  \\
  \hline%
  \tend
  \end{tabular}
  \caption{\footnotesize \label{tab:thcouplings} \textrm{95\% confidence intervals for the anomalous couplings obtained by marginalization of the two-parameter plots in Fig.~\ref{fig:limits1}.} when taken to be independent.}
  \end{table}

To put our result in perspective, we can compare it to the only
theoretical estimate~\cite{Rao:2020hel} which is based, as ours, on observables derived from the polarizations (but not the entanglement). They find
\begin{equation}
  a_Z = 6.88 \times 10^{-3}\,, \quad \widetilde{a}_Z = 9.53 \times 10^{-3} 
\end{equation} 
at 1$\sigma$, in the linear approximation, and for a luminosity of 1 ab$^{-1}$,
whereas, for HiLumi (3 ab$^{-1}$), we find (95\% CL)
\begin{equation}
  a_Z = 4.0 \times 10^{-4}\,, \quad \widetilde{a}_Z = 8.0 \times 10^{-4}  \, . \label{ZZlimits}
\end{equation} 
Beside the different statistical distribution and luminosities used, our result  is stronger than the former limits  obtained for the associated $H Z$ production process. The limits in \eq{ZZlimits} are so stringent to be comparable with projected bounds from future lepton colliders~\cite{Han:2000mi,Craig:2015wwr,Sharma:2022epc} based on classical spin correlations and cross sections.

Although the use of entanglement does seem to strengthen the constrains on the anomalous couplings, we must include the effect of the backgrounds for a more realistic estimate of its power. 
\subsection{Including the background}

The estimate of the dominant $W$ plus jets background for the $H\to WW^*$  case  is currently affected by a rather large uncertainty and its size is larger than the signal. Although we could  in principle  apply our method, the computation of this background is much more involved and it will require a dedicated simulation, which is outside  the scope of the present analysis.

In the following we focus on the $ZZ^*$ decay that achieves higher sensitivity to the anomalous couplings because of a better signal-to-background ratio than the $WW^*$ case.

The irreducible background for the $H\to Z\ell^+\ell^-$ signal is rather small and dominated by the electroweak process $p p \to  ZZ/Z\gamma \to  4\ell$. The sum of the signal and background, being induced by different production channels, is characterized by a mixture of different states. We introduce this mixture by writing the density matrices for the processes as 
\begin{align}
    \rho_{\mbox{\tiny ZZ}} = \alpha \rho_{\mbox{\tiny H $\to$ ZZ}}+ (1-\alpha) \rho_{\mbox{\tiny BCKG}}   \, ,  
\label{BCKG}
\end{align}
with $0\leq \alpha \leq 1$ parametrizing the expected signal-to-background ratio $S/(S+B)$. 
The density matrix  $\rho_{\mbox{\tiny BCKG}}$ in \eq{BCKG} is given
by the electroweak $pp \to ZZ$ process. We take $\alpha=0.8$, corresponding to a background which is about 4 times smaller than the signal, which is the case at the Higgs peak~\cite{CMS:2021ugl}.

\begin{figure}[h!]
\begin{center} 
\includegraphics[width=3in]{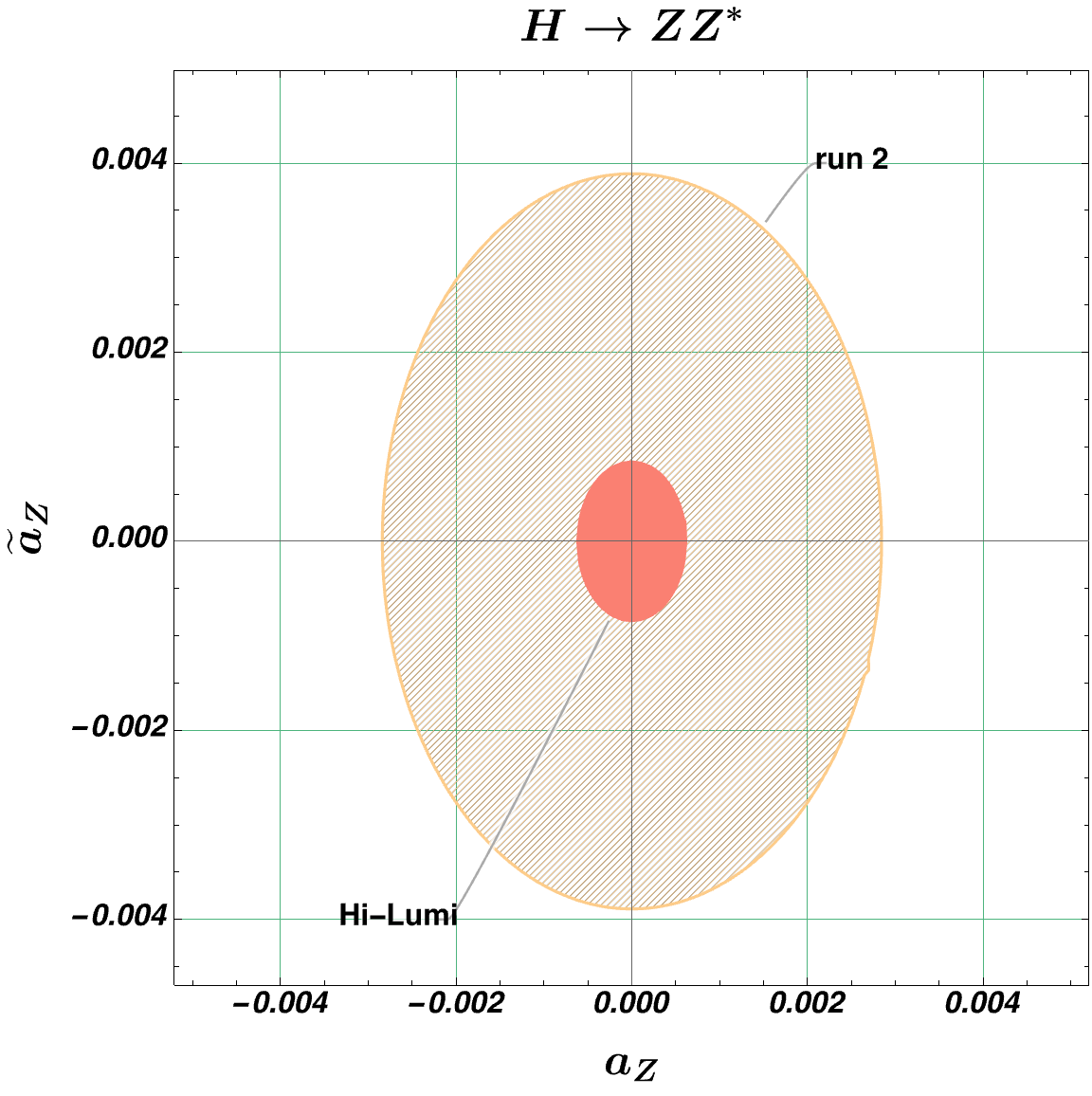}
\caption{\footnotesize The same limits as in Fig.~\ref{fig:limits1} (LHC run2 and HiLumi projections) as obtained for the $ZZ^*$ channel after the inclusion of background. The observables $\cmb$ and $\mathscr{C}_{odd}$ are used in this case. 
\label{fig:bckg} 
}
\end{center}
\end{figure}

Once the background is included, we can no longer use the entropy to measure the entanglement as the produced bipartite state is not pure. We use the observable $\cmb$ instead, which still tracks the entanglement, decreasesing as the latter weakens. The observable $\mathscr{C}_{odd}$ does not receive any contribution from the backgrounds, which are CP even. 

\begin{table}[t]
\tablestyle[sansboldbw]
\begin{tabular}{*{3}{p{0.2\textwidth}}}
\theadstart
    \thead LHC &\thead  run2&
    \thead  HiLumi  \\
\tbody
   & $ |a_{Z}| \leq 0.0028 $   & $|a_{Z}|\leq 0.00062$ \\
  &   $|\widetilde a_{Z}|\leq 0.0039$   & $|\widetilde a_{Z}|\leq 0.00086$  \\
   \hline%
  \tend
\end{tabular}
\caption{\footnotesize \label{tab:couplings} \textrm{95\% confidence intervals for the anomalous couplings obtained by marginalization of the two-parameter plots in Fig.~\ref{fig:bckg}.} when taken to be independent.}
\end{table}

\section{Outlook}
 
{\versal We have outlined a strategy} to improve the current constraints on the anomalous couplings of the Higgs boson to the weak gauge bosons by means of the quantum tomography of the Higgs boson decay.

When comparing the limits found for the $H\to ZZ^*$ process in presence of background processes to those reported by the CMS collaboration one has to write them, in terms of the parameters $f_{g2}$ and $f_{g3}$ introduced in~\cite{Anderson:2013afp}, as
\be
 f_{g2}= \frac{\sigma_{2}}{\sigma} \, \left| a_V \right|^{2}\, ,\quad \text{and} \quad f_{g3}= \frac{\sigma_{3}}{\sigma} \, \left| \widetilde a_V \right|^{2}\, ,
\ee
where we take all anomalous coupling to be real and $\sigma_{i}$ is the cross section in which only the corresponding coupling is included, $\sigma$ the total cross section with all couplings included. Taking the values in Tab.~\ref{tab:couplings}, we have for run2 at the LHC
\be
  f_{g2}^Z<7.8\times 10^{-6}\, , \quad f_{g3}^Z <1.5 \times 10^{-5}\, , \label{us}
\ee
which can be compared to the best current experimental bounds from the CMS collaboration~\cite{CMS:2019ekd}:
\be
   f_{g2}^V< 3.4 \times 10^{-3}\, , \quad f_{g3}^V< 1.4  \times 10^{-2} \, , \label{CMS}
\ee
obtained by using a combination of cross sections, including production ones,  for processes where the $HZZ$ vertices are identified. A comparison of the limits thus found with those of the experimental collaboration shows how  well the proposed set of observables perform. In comparing, one must bear in mind that our limits come from a single process while those with which we  compare come from the simultaneous use of more cross sections. On the other hand, our limits are mostly idealized whereas those from the experimental collaborations come with a full estimate of the statistical and systematic uncertainties as well as the actual backgrounds.

It goes without saying that all the limits we quote depend on the uncertainty in the analysis (about which we  have made an educated guess)  and that they can become stronger or weaker depending on how well (or badly) the actual physical analysis will turn out to be. In this respect, the main purpose of the present analysis is to highlight the potential of the  new observables in constraining the anomalous couplings.
In this light, we proved that their inclusion in routine analyses could lead to more stringent limits from global fits, which are currently based only on cross sections.

\twocolumn  
\section*{Acknowledgements}
\small
L.M. is supported by the Estonian Research Council grant PRG356.

\bibliographystyle{JHEP}   
\bibliography{HVV.bib} 

\end{document}